\documentclass[prl,twocolumn,superscriptaddress]{revtex4-1}
\usepackage{}
\usepackage{amssymb}
\usepackage{amsfonts}
\usepackage{bbm}
\usepackage{mathrsfs}
\usepackage{graphics,graphicx,epsfig,bm,amsmath,amsthm,amssymb}
\usepackage{bm}
\usepackage{bbm}
\usepackage{longtable}
\usepackage{multirow}
\usepackage{array}
\usepackage{color}
\usepackage{diagbox}
\usepackage{amsmath}
\usepackage{upgreek}
\usepackage{multirow}
\usepackage{threeparttable}
\usepackage{braket}
\usepackage{mathtools}
\usepackage[usenames,dvipsnames]{xcolor}

\usepackage{float}
\usepackage[a4paper,colorlinks=true,
linkcolor=blue,citecolor=blue,
pdfauthor={ },
pdftitle={ },
pdfsubject={ },
pdfkeywords={ }]{hyperref}

\newcommand{\tabincell}[2]{\begin{tabular}{@{}#1@{}}#2\end{tabular}}
\begin{document}

	\title{Non-Hermitian non-Abelian topological transition in the S=1 electron spin system of a nitrogen vacancy centre in diamond}
	\affiliation{CAS Key Laboratory of Microscale Magnetic Resonance and School of Physical Sciences, University of Science and Technology of China, Hefei 230026, China}
	\affiliation{Hefei National Laboratory, University of Science and Technology of China, Hefei 230088, China}
	\affiliation{Anhui Province Key Laboratory of Scientific Instrument Development and Application, University of Science and Technology of China, Hefei 230026, China}
	\affiliation{Beijing National Laboratory for Condensed Matter Physics, Institute of Physics, Chinese Academy of Sciences, Beijing 100190, China}
	\affiliation{School of Physical Sciences, University of Chinese Academy of Sciences, Beijing 100049, China}
	\affiliation{Institute of Quantum Sensing and School of Physics, Zhejiang University, Hangzhou 310027, China}
	
	\author{Yunhan Wang}
	\thanks{These authors contributed equally to this work.}
	\affiliation{CAS Key Laboratory of Microscale Magnetic Resonance and School of Physical Sciences, University of Science and Technology of China, Hefei 230026, China}
	\affiliation{Hefei National Laboratory, University of Science and Technology of China, Hefei 230088, China}
	\affiliation{Anhui Province Key Laboratory of Scientific Instrument Development and Application, University of Science and Technology of China, Hefei 230026, China}
	
	\author{Yang Wu}
	\thanks{These authors contributed equally to this work.}
	\affiliation{CAS Key Laboratory of Microscale Magnetic Resonance and School of Physical Sciences, University of Science and Technology of China, Hefei 230026, China}
	\affiliation{Anhui Province Key Laboratory of Scientific Instrument Development and Application, University of Science and Technology of China, Hefei 230026, China}

	\author{Xiangyu Ye}
	\affiliation{CAS Key Laboratory of Microscale Magnetic Resonance and School of Physical Sciences, University of Science and Technology of China, Hefei 230026, China}
	\affiliation{Anhui Province Key Laboratory of Scientific Instrument Development and Application, University of Science and Technology of China, Hefei 230026, China}
	
	\author{Chang-Kui Duan}
	\affiliation{CAS Key Laboratory of Microscale Magnetic Resonance and School of Physical Sciences, University of Science and Technology of China, Hefei 230026, China}
	\affiliation{Hefei National Laboratory, University of Science and Technology of China, Hefei 230088, China}
	\affiliation{Anhui Province Key Laboratory of Scientific Instrument Development and Application, University of Science and Technology of China, Hefei 230026, China}
	
	\author{Ya Wang}
	\affiliation{CAS Key Laboratory of Microscale Magnetic Resonance and School of Physical Sciences, University of Science and Technology of China, Hefei 230026, China}
	\affiliation{Hefei National Laboratory, University of Science and Technology of China, Hefei 230088, China}
	\affiliation{Anhui Province Key Laboratory of Scientific Instrument Development and Application, University of Science and Technology of China, Hefei 230026, China}
	
	\author{Haiping Hu}
	\email{hhu@iphy.ac.cn}
	\affiliation{Beijing National Laboratory for Condensed Matter Physics, Institute of Physics, Chinese Academy of Sciences, Beijing 100190, China}
	\affiliation{School of Physical Sciences, University of Chinese Academy of Sciences, Beijing 100049, China}
	
	\author{Xing Rong}
	\email{xrong@ustc.edu.cn}
	\affiliation{CAS Key Laboratory of Microscale Magnetic Resonance and School of Physical Sciences, University of Science and Technology of China, Hefei 230026, China}
	\affiliation{Hefei National Laboratory, University of Science and Technology of China, Hefei 230088, China}
	\affiliation{Anhui Province Key Laboratory of Scientific Instrument Development and Application, University of Science and Technology of China, Hefei 230026, China}

	\author{Jiangfeng Du}
	\email{djf@ustc.edu.cn}
	\affiliation{CAS Key Laboratory of Microscale Magnetic Resonance and School of Physical Sciences, University of Science and Technology of China, Hefei 230026, China}
	\affiliation{Hefei National Laboratory, University of Science and Technology of China, Hefei 230088, China}
	\affiliation{Anhui Province Key Laboratory of Scientific Instrument Development and Application, University of Science and Technology of China, Hefei 230026, China}
	\affiliation{Institute of Quantum Sensing and School of Physics, Zhejiang University, Hangzhou 310027, China}

	\begin{abstract}
	Topological phases and transitions are of fundamental importance in physics, which provide a deep insight into the understanding of materials. 
	Recently, non-Abelian topological transitions have been investigated in Hermitian systems, revealing important topological features. 
	With non-Hermiticity introduced, non-Hermitian non-Abelian topological transitions bring about more intriguing topological features, yet has not been experimentally explored. In this work, we report the observation of the non-Hermitian non-Abelian topological transition at the atomic scale utilizing a nitrogen-vacancy center in diamond.
	While the well-established topological numbers, failed to recognize this transition, we successfully characterized such a transition with the measurement of the complex eigenvalue braids. We obtained the braid invariants from the measured relative phases between eigenvalues. The observed change in braid invariants provides a clear signature of the non-Abelian topological transition.
	Furthermore, we experimentally revealed an intriguing consequence of this transition, which is the creation of a third-order exceptional point through the collision of two second-order exceptional points with opposite charges.
	Our experimental findings shed light on the abundant non-Abelian topological phenomena involving non-Hermiticity, and provide insights into manipulating the spectral topology in atomic scale systems to achieve exotic functionalities arising from non-Abelian band braiding.
	\end{abstract}

	\maketitle
The Aharonov-Bohm effect\cite{ABeff}, quantum Hall effect\cite{qHall} and topological insulators\cite{insulator1,insulator2}, are all tied to a topological origin. Global properties of a system are often governed by topological invariants, which stay intact under continuous deformations with small variations of system parameters. When the system's global topological properties change, which is caused by gap opening or closing, topological transitions occur, marked by changes in the topological invariants\cite{transition_Hermi1,transition_Hermi2,transition_Hermi3}. The form of topological invariants reflects its topological origin. For example, integer or $\mathbb{Z}_2$ values of topological invariants were Abelian since they behave like numbers. Some topologies were not described by numbers, but non-commuting operators, where the topological invariants became non-Abelian. These intriguing topologies gave rise to unconventional topological phases\cite{nattH1,nattH2,nattH3,nattH4,nattH5,nattH6} and non-Abelian topological transitions beyond traditional classifications based on Altland-Zirnbauer class\cite{A,B,C}, where different symmetry conditions were considered.

With the introduction of non-Hermiticity, band degeneracies known as exceptional points (EPs)\cite{ep1,ep2} came into play for systems described by non-Hermitian (NH) Hamiltonians\cite{revmodphys,naturereview}. Unlike Hermitian systems with real eigenvalues, the EPs resulted in non-trivial spectral topology\cite{nh_spectra1,nh_spectra2,nh_spectra3,nh_spectra4,nh_spectra5} of NH systems with extraordinary properties and applications\cite{winding1,closegap1,App1,App2,App3,App4,App5,App6,Chiral1,Chiral2,Chiral3,unidirec1,unidirec2,braiding1,HarrisNature}. 
Around EPs, the eigenvalues might get tangled with each other in the complex plane, forming eigenvalue braids\cite{nh_spectra1,braidgroup2}. 
The NH non-Abelian topological transition, arising from dynamical interplay between multiple types of EPs, defines a distinct category of topological phenomena.
These transitions provide guidance on achieving desired functionalities for classical\cite{function1,function2} and quantum systems. 
For instance, through these transitions, manipulating the spectral topology\cite{manipulating2} and constructing high-order singularities\cite{highorderep1} can achieve advanced control schemes for atomic scale systems, making multimode switching possible\cite{DDEP}. 
While complex energy braiding has recently been demonstrated\cite{braiding1,HarrisNature}, experimental observation of NH non-Abelian topological transitions remains elusive.

In this work, we observe NH non-Abelian topological transitions at the single-spin scale. 
The NH system is engineered in the single electron spin of a nitrogen-vacancy (NV) centre in diamond, by a dilation method with a nearby nuclear spin. 
The measured variation of the relative phases between energy bands successfully capture the eigenvalue braids, distinguishing different braid invariants.
We demonstrate that the braid invariant of eigenvalues changes from identity to a non-trivial element of the braid group when the movement of EPs induces band closing in a specific region. The change in braid invariants successfully characterizes the non-Abelian topological transition, while the conventional topological invariants like discriminant numbers remain unchanged and fail to identify this transition. Furthermore, we present an intriguing phenomenon resulting from this transition: two pairwise-created second-order EPs (EP2s) of opposite charges do not annihilate but produce a third-order EP (EP3) after the non-Abelian topological transition.
	
	\begin{figure}[http]
		\centering
		\includegraphics[width=1\columnwidth]{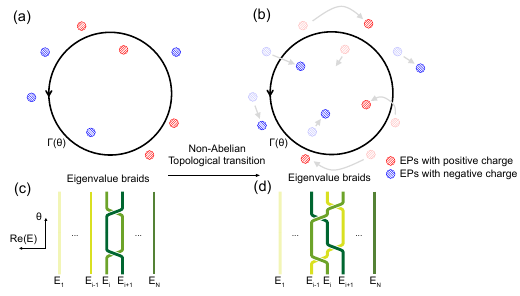}
		\caption{Schematics of the moving EPs in the parameter space and the braid invariants.
			(a) Initial positions of the EPs in the parameter space and the corresponding eigenvalue braids on $\Gamma$. The red and blue circles correspond to EPs with positive and negative charges, respectively. The black circle mark the trajectory $\Gamma$. (b) The positions of EPs when the control parameter changes. Gray lines with arrows in (b) are the trajectories of EPs. (c,d) The schematics of eigenvalue braids along $\Gamma$. The bands are labeled according to its real parts. After outer EPs enter $\Gamma$, while the total charge within $\Gamma$ is still zero, the eigenvalue braids change.
		}
		\label{Fig1}
	\end{figure}
	
\section{Non-Hermitian non-Abelian topological transition}
We investigate the change in spectral topology in a two-dimensional parameter space when the position of EPs varies according to a control parameter. As shown in Fig.\ref{Fig1}, the focused region is enclosed by an oriented curve $\Gamma$. The spectral topology in this region is closely related to the EPs enclosed by $\Gamma$, since the complex bands get tangled around EPs.
Topological charges, which are positive or negative, could be assigned to EPs\cite{EPannihilate2,EPannihilate} (red or blue circles in Fig.\ref{Fig1}(a,b)).
Akin to the case of Ampere's Law, the global spectral topology of the region is captured by the information on its boundary $\Gamma$. For example, the total charge enclosed by $\Gamma$ equals the discriminant number along $\Gamma$.
Inflows and outflows of EPs passing through $\Gamma$ induce topological transitions, which are indexed by the change in global topological invariants such as the total charge within $\Gamma$. 
For the schematic in Fig.\ref{Fig1}(a,b), the total charge remains zero, since the numbers of positive and negative charges are equal. However, this does not mean that the cases in Fig.\ref{Fig1}(a,b) are topologically equivalent. 
This is because there are different types of EPs, even if they possess the same charges. 
For example, in three-band systems, two EPs could couple the bands labeled by 1 and 2, 2 and 3, respectively. 
They yield different braids of eigenvalues (Supplementary Materials, section 6). 
In order to characterize the transition, the dynamical interplay between multiple EPs should be considered.
Tracing along $\Gamma$, the real parts of eigenvalues may cross each other, as shown in the braiding graphs in Fig.\ref{Fig1}(c,d), where the eigenvalues are flattened and projected onto the real axis.  
Each crossing can be described by the element $\sigma_{i(i+1)}$, which is the generator of the braid group\cite{braidgroup1,braidgroup2}, $B(N)$ ($N$ is the number of eigenvalues). 
The physical meaning of $\sigma_{i(i+1)}$ ($\sigma_{i(i+1)}^{-1}$) is the band with $(i+1)^{\text{th}}$ largest real part crosses the band with $i^{\text{th}}$ largest real part from above (below). 
A single operator $\sigma_{i(i+1)}$ is similar to the vorticity between two bands\cite{braiding1}, as $\sigma_{i(i+1)}^n$ means the two bands are tangled $n$ rounds. 
Different operators can exhibit non-commuting relation. 
Since after the action of $\sigma_{i(i+1)}$, the band indices $i$ and $i+1$ are interchanged, the relation between $\sigma_{i(i+1)}$ and $\sigma_{(i-1)i}$ is $\sigma_{i(i+1)}\sigma_{(i-1)i}\sigma_{i(i+1)}=\sigma_{(i-1)i}\sigma_{i(i+1)}\sigma_{(i-1)i}$. 
This reflects the topological origin from the braiding behaviors of eigenvalues. 
The eigenvalue braid after one cycle around the trajectory corresponds to an element $b$ in $B(N)$, which is obtained by multiplying each generator following the trajectory. 
For example, the elements are $b=\sigma_{i(i+1)}\sigma_{i(i+1)}^{-1}=\mathbb{I}$ and $b=\sigma_{(i-1)i}\sigma_{i(i+1)}\sigma_{(i-1)i}^{-1}\sigma_{i(i+1)}^{-1}$ for Fig.\ref{Fig1}(c) and (d), respectively. 
If no EPs enter or exit the curve, the element remains unchanged (up to conjugate equivalence) according to the non-Abelian conservation rule\cite{function2}. 
Thus, the element $b$ is a well-defined braid-valued topological invariant. 
When there are inflows or outflows of EPs, non-Abelian topological transitions occur.
The change in braid invariants between Fig.\ref{Fig1}(d) and Fig.\ref{Fig1}(c) characterizes such transition, while the total charge within $\Gamma$ remains conserved.
	
	\begin{figure}[http]
		\centering
		\includegraphics[width=1\columnwidth]{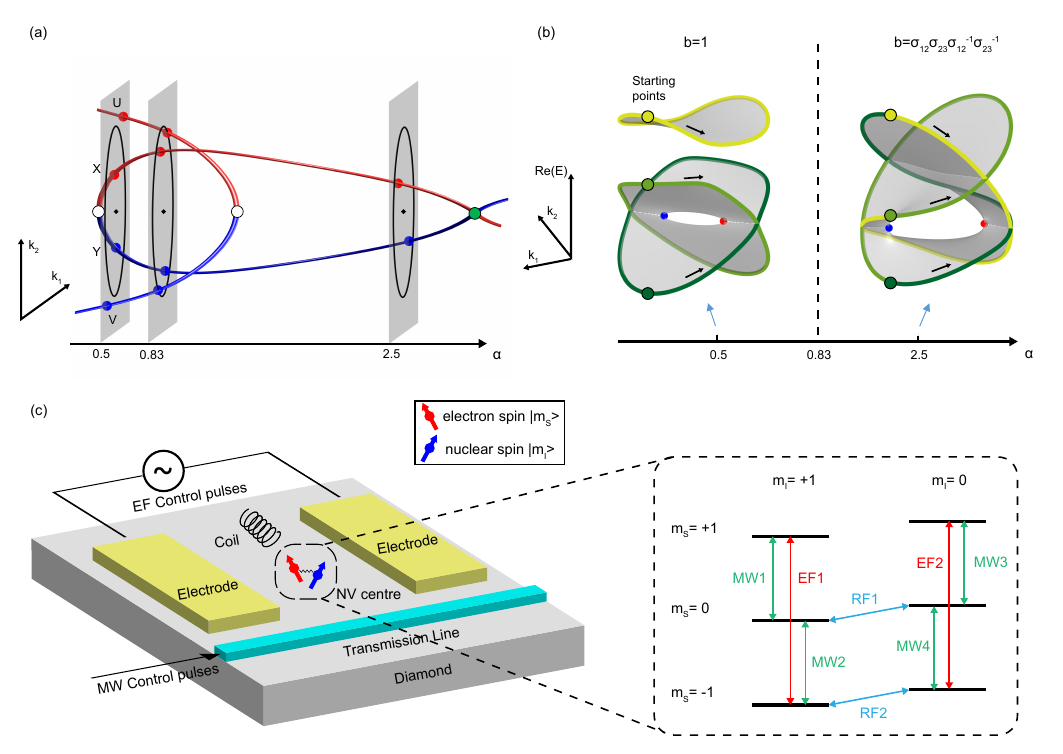}
		\caption{The model of the NH Hamiltonian and experimental system of an NV centre.
			(a) Paths of EPs of the Hamiltonian described by Eq.\ref{Model} with respect to the parameter $\alpha$. The shadowed planes mark the sections at different values of $\alpha$, and the black curves are the trajectories $\Gamma$ at different sections. The red and blue lines are the trajectories of EPs with positive and negative charges, respectively. Red and blue points mark the positions of the EPs named by $X,Y,U,V$. The white circles represent EP2s and green circle represents a EP3. (b) Transition in spectral topology on the region enclosed by $\Gamma$.
			Gray surfaces are the Riemann sheets of the real part of eigenvalues. Red (blue) points correspond to the EP $X(Y)$. Eigenvalues on $\Gamma$ are marked as the colored lines on the edge of the surfaces. The arrows mark the orientation of $\Gamma$, and starting points correspond to $\theta=0$. 
			(c) Experimental system of an NV centre. Microwave (MW) and elecric field (EF) pulses are applied by transmission line and electrodes fabricated on diamond surface, respectively. The radio-frequency (RF) pulses are delivered via a home-build coil. The six energy levels of the ground state of the NV centre are used in the experiment with $m_S$ and $m_I$. The RF pulses (blue arrows) are applied for the initial state preparation of the nuclear spin. MW (green arrows) and EF pulses (red arrows) are implemented to realize the NH Hamiltonian.
		}
		\label{Fig2}
	\end{figure}

As an example, we consider the following form of an NH Hamiltonian:
\begin{equation}
	\begin{aligned}
		H_\alpha(k_1,k_2) &= \sqrt{2}[(i\frac{\alpha+1}{4}-k_1)S_z-S_x]\\
		&+\lbrace ik_2-[1-(\alpha-2)^2]\rbrace\left[\begin{array}{ccc}
			0 & 0 & 0\\
			0 & 1 & 0\\
			0 & 0 & 0
		\end{array}\right],
	\end{aligned}
	\label{Model}
\end{equation}
where $k_1,k_2,\alpha$ are parameters of the Hamiltonian, $S_{x,y,z}$ are spin-1 operators. 
This three-band model is chosen to realize the EP-trajectories in Fig.\ref{Fig2}(a), which enable the occurrence of the NH non-Abelian topological transition. 
We focus on the region of a disk centered at the origin (black dot in Fig.\ref{Fig2}(a)) in the $k_1$-$k_2$ plane and investigate the dependence of the spectral topology on the parameter $\alpha$ varying from 0 to 3. 
The boundary $\Gamma$ is parameterized by $\theta$ as $\Gamma(\theta)=(r\cos(\theta),r\sin(\theta))$ (black curves in Fig.\ref{Fig2}(a)).
The positions of EPs can be found by solving $\Delta_\alpha(k_{1,EP},k_{2,EP})=0$, where $\Delta_\alpha$ is the discriminant of the characteristic polynomial of $H_\alpha$ (Supplementary Materials, section 3). 
When $\alpha$ varies, the EPs not only move in the parameter space, but also can be created or annihilated. 
As shown in Fig.\ref{Fig2}(a), a pair of EPs (labeled by $X,Y$) with opposite charges (discriminant number $\pm1$) is initially created from the vacuum at the origin and then splits, with the EP carrying positive (negative) charge represented by the red (blue) line. 
Another two EPs, labeled by $U,V$ with opposite charges in Fig.\ref{Fig2}(a), are outside $\Gamma$ at the beginning $\alpha=0$. 
Before $U,V$ crossing $\Gamma$, the upper band is gapped from the lower two bands, and each eigenvalue at the starting points along $\Gamma$ returns to itself, as shown in left panel of Fig.\ref{Fig2}(b) for the typical scenario of $\alpha=0.5$.
The corresponding braid invariant is $b=\mathbb{I}$. 
As $\alpha$ increases, EPs $U,V$ move towards the origin, pass between EPs $X$ and $Y$, and finally annihilate before the coalescence of $X,Y$. 
As will be shown in the following, the spectral topology enclosed by $\Gamma$ changes when $U,V$ enter $\Gamma$. 
There is a point where they flow into the region enclosed by $\Gamma$, at which the distance between EPs and the origin satisfies $r_{U,V}(\alpha_0)^2=r^2$.
The EPs $U,V$ flow in simultaneously since their positions are symmetric with respect to the origin.
In Fig.\ref{Fig2}(a,b), $r$ is chosen as $1.4$ and the non-Abelian topological transition occurs at $\alpha_0=0.83$, where $U,V$ flow in $\Gamma$.
Following the transition, the spectral topology of the enclosed region undergoes a transformation, resulting in the closure of the gap between bands 1 and 2. 
The change in the spectral topology is also manifested in the eigenvalue braid. 
After the transition, each band can only return to itself after three cycles along $\Gamma$ as shown in right panel of Fig.\ref{Fig2}(b) for the typical scenario of $\alpha=2.5$.
The associated braid invariant is $b=\sigma_{12}\sigma_{23}\sigma_{12}^{-1}\sigma_{23}^{-1}$. 
Despite non-intersection between the EPs $U,V$ and $X,Y$ at any $\alpha$, which preserves both individual EP charges and a zero total charge within $\Gamma$, the EPs $U,V$ still alter the properties of EPs $X,Y$. 
A common understanding predicted that two EPs with opposite charges annihilate upon collision\cite{EPannihilate2,EPannihilate}. 
However, the scenario here demonstrates the emergence of a EP3 instead (green circle in Fig.\ref{Fig2}(a)). 
This change of the EPs' properties is successfully captured by the change of braid invariant from the trivial case $b=\mathbb{I}$ to a nontrivial operator $b=\sigma_{12}\sigma_{23}\sigma_{12}^{-1}\sigma_{23}^{-1}$, while the charge fails to identify this process.

\section{Hamiltonian construction in an NV centre spin system}
The NV centre in diamond was utilized to investigate the NH non-Abelian topological transition (Fig.~\ref{Fig2}c and Methods). 
The NV centre is an atomic-scale defect in diamond, comprising a vacancy and a neighboring nitrogen atom. 
The ground state of the NV centre can be treated as two coupled spin-1 system $\ket{m_S=0,\pm1}_e\ket{m_I=0,\pm1}_n$, with one describing the electron spin and the other for the nuclear spin. 
Similar to the method described in literatures\cite{HarrisNature,braiding1,nh_spectra1}, we characterize the braid invariants through the pattern of complex eigenvalues along a closed loop.
With prior knowledge that the Hamiltonian takes the form given in Eq.\ref{Model}, the eigenvalues can be obtained by tomography on the eigenstates (Methods).
The eigenstates are prepared through the evolution under NH Hamiltonians with different parameters along the loop.
The NH Hamiltonians are realized based on the dilation method\cite{dilation} in the following procedure (Methods). 
The subspace spanned by $\ket{m_S=0,\pm1}_e\ket{m_I=0,1}_n$, is utilized to construct the dilated Hamiltonian. 
The initial state is prepared to the form $\ket{\Psi(0)}=\ket{\psi(0)}_e(\ket{-}_n+\eta_0\ket{+}_n)$ by microwave and radio-frequency pulses, where $\ket{\pm}_n$ are the eigenstates of $\sigma_y$. 
Then the state is evolved under the dilated Hamiltonian. 
The dilated Hamiltonian has the form $H_{tot}(t) = \Xi(t)\otimes\ket{1}_n \prescript{}{n}{}\bra{1}+\Lambda(t)\otimes\ket{0}_n\prescript{}{n}{}\bra{0}$, where $\Xi(t)$ and $\Lambda(t)$ are determined by $H_\alpha(k_1,k_2)$ (Supplementary Materials, section 1). 
The diagonal elements of $H_{tot}$ are realized by an appropriate interaction picture. 
The non-diagonal elements of $H_{tot}$ are realized by control pulses with time-dependent amplitudes, frequencies and phases corresponding to transitions among $\ket{0}_e\ket{m_I}_n$, $\ket{1}_e\ket{m_I}_n$ and $\ket{-1}_e\ket{m_I}_n$. 
The state evolved under $H_{tot}$ for time $t$ is $\ket{\Psi(t)}=\ket{\psi(t)}_e\ket{-}_n+\eta(t)\ket{\psi(t)}_e\ket{+}_n$. 
Here, in the $\ket{-}_n$ subspace, $\ket{\psi(t)}_e$ effectively evolves under the NH Hamiltonian up to a normalization factor. 
When the evolution time $t$ is long enough, the $\ket{\psi(t)}_e$ approaches an eigenstate (Supplementary Materials, section 5). 
Following the operation rotating $\ket{\pm}_n$ to $\ket{0,1}_n$, populations of each level are measured as $P_{\ket{m_S}_e\ket{m_I}_n}$.
Finally, the population of the eigenstate within the target subspace can be obtained as $P_{\ket{m_S}_e} = P_{\ket{m_S}_e\ket{1}_n}/(\sum_{i=1}^{3}P_{\ket{i}_e\ket{1}_n})$\cite{tomo}. 

\section{Topological transition}
\begin{figure}[http]
		\centering
		\includegraphics[width=1\columnwidth]{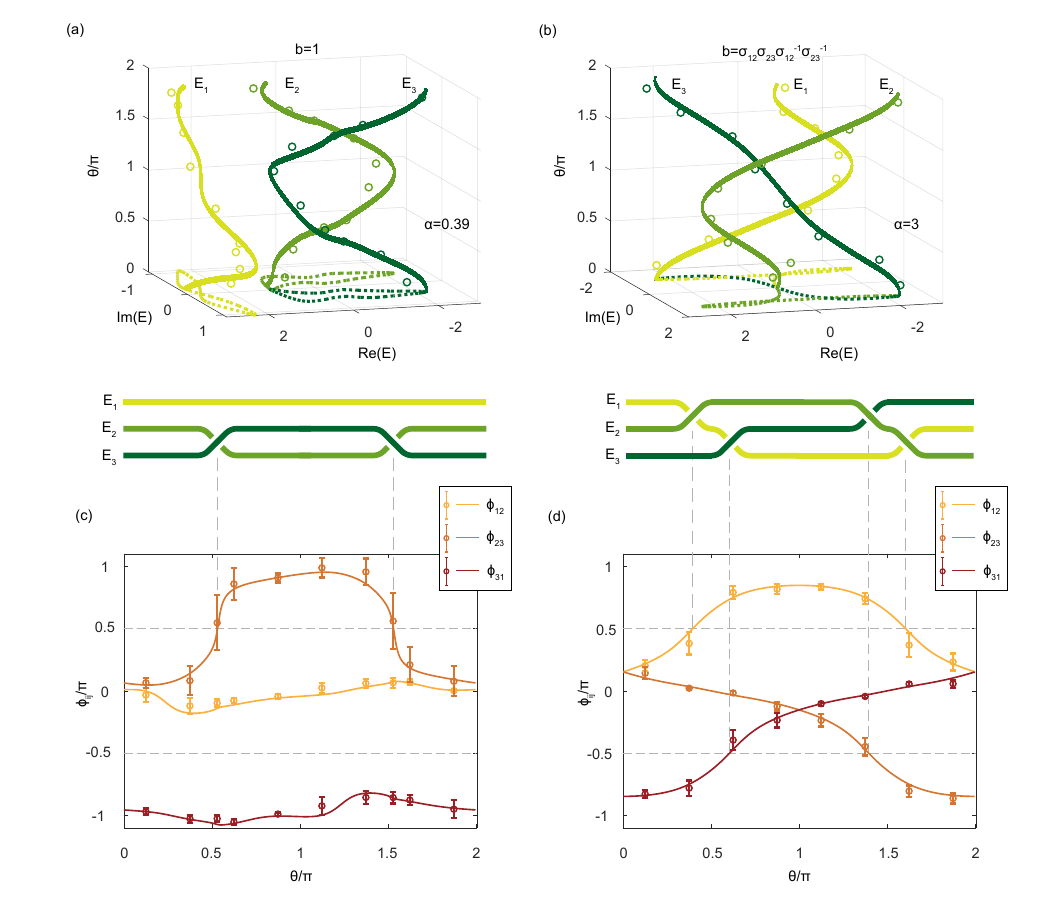}
		\caption{Observation of the non-Abelian topological transition by characterizing the change in eigenvalue braids.
			(a,b) The eigenvalue braids along the trajectory $\Gamma$ corresponding before ($\alpha=0.39$) and after ($\alpha=3$) the non-Abelian topological transition. Solid lines are numerical predictions and circles are the experimental results. The dashed lines are projection of the solid lines to the complex plane. The schematics of corresponding eigenvalue braids are given in the lower panel. (c,d) The variation of relative phases $\phi_{ij}$ along $\Gamma$. In (c), all of the three $\phi_{ij}$ go back to the same value at $\theta=0$ as $\theta$ goes to $2\pi$. In (d), the final values of $\phi_{31}$ and $\phi_{23}$ exchange each other. The solid lines are numerical predictions and the circles with error bars of the corresponding colors are the experiment results. All errors shown are one standard derivation with 0.6 million averages.
		}
		\label{Fig3}
	\end{figure}

The non-Abelian topological transition is manifested by the change in the eigenvalue braids on the trajectory $\Gamma$ as $\alpha$ varies from below to above the transition point. 
The radius of $\Gamma$ is $r=1.4$ and the transition point is $\alpha_0=0.83$. 
In our experiment $\alpha=0.39$ and $3$ are chosen to demonstrate the change in the eigenvalue braids. 
In order to clearly characterize the eigenvalue braids, the relative phases between the bands, defined as $\phi_{ij}(\theta)=-\text{arg}(E_i(\theta)-E_j(\theta))$\cite{phase} are measured. 
Here, the measured eigenvalues are grouped according to the fact that the eigenvalues should vary continuously with no crossings when the parameter $\theta$ varies. 
According to the definition of $\phi_{ij}$, the crossing of the real part of the eigenvalues occurs when $\phi_{ij}(\theta) = \pm\pi/2$. 
Fig.\ref{Fig3} shows the results of eigenvalue braids as well as the relative phases corresponding to $\Gamma$ at $\alpha = 0.39$ and $3$.
When $\alpha<\alpha_0$, i.e., before the EPs $U,V$ enter $\Gamma$, as shown in Fig.\ref{Fig3}(a), there are two crossings between the green and dark green bands. And the yellow band does not cross with the other two bands. Correspondingly, relative phases in Fig.\ref{Fig3}(c) show that $\phi_{23}(\theta)$ crosses $\pi/2$ twice, while $\phi_{12}(\theta)$ and $\phi_{31}(\theta)$ do not cross the dashed gray lines representing $\pm\pi/2$. The elements representing these two crossings can be determined as $\sigma_{23}$ and $\sigma_{23}^{-1}$. Thus, the braid invariant is $b=\sigma_{23}\sigma_{23}^{-1}=\mathbb{I}$, which is the trivial case (Methods).
For the case $\alpha>\alpha_0$, i.e., $U,V$ are inside $\Gamma$. 
Fig.\ref{Fig3}(b) shows that the eigenvalues are braided in a non-trivial way. 
Tracing along $\Gamma$, the eigenvalues at the beginning and end points permute. 
The movement of EPs $U,V$ flowing into $\Gamma$ is manifested by the changes in the crossings between $\phi_{ij}$ and $\pm\pi/2$. 
Compared to the trivial case, there are four crossings shown in Fig.\ref{Fig3}(d) determined to be $\sigma_{12}$, $\sigma_{23}$, $\sigma_{12}^{-1}$ and $\sigma_{23}^{-1}$, respectively, as shown in the schematic of braids in Fig.\ref{Fig3}(b).
The braid invariant is then $b=\sigma_{12}\sigma_{23}\sigma_{12}^{-1}\sigma_{23}^{-1}$. 
Since $\sigma_{23}$ and $\sigma_{12}$ do not commute, the braid invariant cannot be reduced to identity. 
The change of braid invariant signifies a non-Abelian topological transition, arising from the dynamical interplay between multiple EPs that defines their non-Abelian nature.
	
\begin{figure}[http]
		\vspace{-0em}
		\centering
		\includegraphics[width=1\columnwidth]{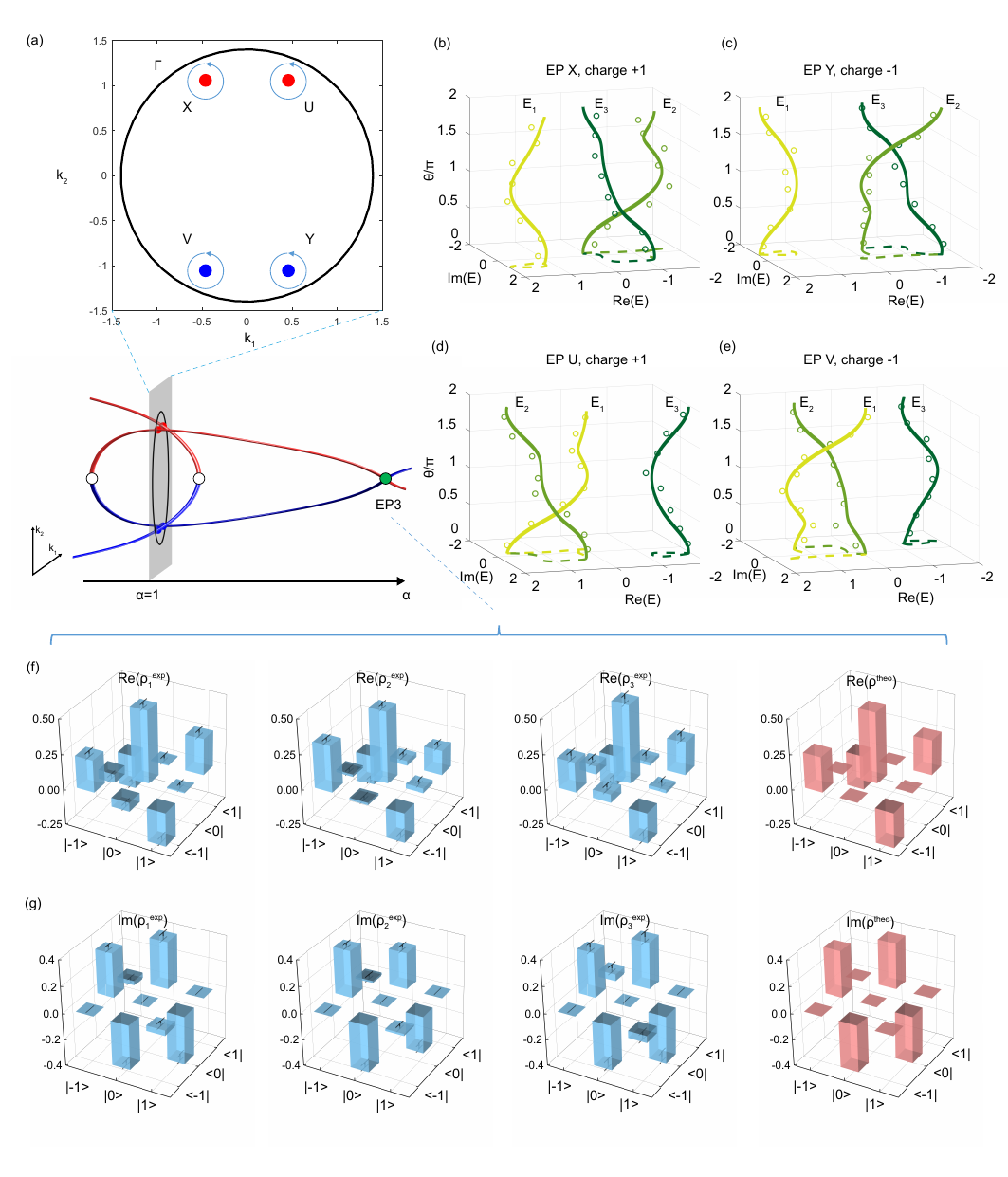}
		\caption{Measuring the charges of different EPs and the third-order EP produced by collision of EPs $X,Y$.
			(a) Local encircling curves of EPs and the trajectories of EPs with respect to $\alpha$. The encircling trajectory is chosen as the counter-clockwise oriented circle centered at the EPs with radius 0.5 in the section $\alpha=1$. (b-e) The local eigenvalue braids corresponding to the EPs $X,Y,U,V$. According to the definition in Eq.\ref{Eq2}, the EPs $X,U$ possess positive charges and the EPs $Y,V$ have negative charges. (f-g) Blue bars with error bars are real (f) and imaginary (g) parts of the measured density matrices $\rho_i^{\rm exp}$ obtained by quantum state tomography at the EP3 ($\alpha=3, (k_1,k_2)=(0,0)$). $\rho^{\rm theo}$ is the density matrix of theoretically predicted eigenstate shown by the red bars. All errors shown are one standard deviation with 0.7 million averages.
		}
		\label{Fig4}
	\end{figure}

The non-Abelian topological transition characterized by the braid invariants cannot be identified by the topological charges like discriminant number. The charge associated with each EP is quantified using the discriminant number, which was defined as\cite{EPannihilate2,EPannihilate}
\begin{equation}
	\nu(Z)=\frac{i}{2\pi}\int_{\gamma(Z)}d\beta\nabla_\beta\lbrace\log \mathrm{Disc}[H(\beta)]\rbrace,
	\label{Eq2}
\end{equation}
where $\gamma(Z)$ (parametrized by $\beta$) is a local encircling curve around a point $Z$, and $\mathrm{Disc}[H] = \prod_{i<j}(E_i-E_j)^2$ is the discriminant of the characteristic polynomial of $H$. The physical meaning of the charge was the summation of vorticities\cite{EPannihilate2,EPannihilate} between any two of the three eigenvalue bands. And the charge is positive (negative) when the $(i+1)^{\rm th}$ band crosses the $i^{\rm th}$ band from above (below). Local encircling curves around EPs are chosen to characterize the charges of the EPs $X,Y,U,V$, as shown in the upper panel in Fig.\ref{Fig4}(a). 
The results of the tangled eigenvalues around EPs in Fig.\ref{Fig4}(b-e) show that the charges are $+1,-1,+1$ and $-1$ for the EPs $X,Y,U$ and $V$, respectively. 
The total charge equaled the summation of all charges carried by the EPs included by $\Gamma$\cite{EPannihilate2,EPannihilate}, which is zero both before and after the non-Abelian topological transition. 
The total charge fails to describe the non-Abelian topological transition because it lacks full band-topology information by omitting the braiding behavior of all three eigenvalue bands.
	
An intriguing consequence arises due to the non-Abelian topological transition. 
Although EPs $X,Y$ are born with opposite charges, after the interplay with $U,V$, they produce an EP3 instead of annihilation at $\alpha=3$, as shown in the lower panel in Fig.\ref{Fig4}(a). 
To observe such a phenomenon, overlaps between the eigenstates at the EPs were measured to characterize order of these EPs\cite{tomo}. 
Here, the overlap between two of the eigenstates is defined by $F_{ij} = [\text{Tr}(\sqrt{\sqrt{\rho_i}\rho_j\sqrt{\rho_i}})]^2$, where $\rho_i$ is the measured density matrix from quantum state tomography (Supplementary Materials, section 5) corresponding to the eigenvalue $E_i, i = 1,2,3$. 
The results are summarized in Table~\ref{Fidelity} and the corresponding quantum state tomography results are shown in Fig.\ref{Fig4}(f-g) and Fig.\ref{ExtFig2}\ref{ExtFig3}\ref{ExtFig4}.
At $\alpha=0.39$, eigenstates $\ket{\psi_2}$ and $\ket{\psi_3}$ are degenerate, while they are almost orthogonal to $\ket{\psi_1}$. 
This shows that the EPs $X,Y$ are created from a EP2. 
Then, the section $\alpha = 1$ is chosen in our experiment to verify the splitting of the two EPs $X,Y$. 
The degeneracies of the eigenstates $\ket{\psi_2}$ and $\ket{\psi_3}$ at the two EPs $X,Y$ confirm the splitting into a pair of EP2s. 
Finally, at $\alpha = 3$ after the transition, all three eigenstates are degenerate and the merging point is an EP3.  
As the merging point of two EPs with opposite charges, the EP3 at $\alpha=3$ has charge zero. 
This can be confirmed from the sum of exponents on braid invariant $\sigma_{12}\sigma_{23}\sigma_{12}^{-1}\sigma_{23}^{-1}$\cite{manipulating2,dfb1}. 
The EP2 at $\alpha = 0.39$, where the two EPs are born, also has charge zero, according to the braid invariant $\sigma_{23}\sigma_{23}^{-1}=\mathbb{I}$. Although the charges of the two important points remain unchanged, the transition changes the properties of EPs, leading to the formation of an EP3. With the braid invariants capturing the global topology, the non-Abelian topological transition provides complete information on the motion of EPs.
\begin{table*}\centering  
	\caption{\textbf{Verify the order of the EPs.} The orders of the EPs are characterized by the overlap, $F_{ij}$ between two of the eigenstates corresponding to the eigenvalues $E_i$ and $E_j$.}
	\textrm{\\}
	\renewcommand{\multirowsetup}{\centering}
	\begin{threeparttable}

		\begin{tabular}{c|c |ccc}  
			\hline\hline
			Parameter $\alpha$ & Locations of EPs in the $k_1$-$k_2$ plane& $F_{12}$ & $F_{13}$ & $F_{23}$ \\ \hline  
			{\tabincell{c}{$\alpha=0.39$}} & $k_1=0$, $k_2=0$ &  0.04(1) & \ 0.04(2) & \bf 0.98(2)\tnote{\rm{[1]}} \\   \hline
			
			\multirow{2}{*}{\tabincell{c}{$\alpha=1$}} & $k_1=0.46$, $k_2=-1.06$ &  0.01(2) &  0.02(2) & \bf 0.98(2) \\
			& $k_1=-0.46$, $k_2=1.06$ & 0.03(2) & 0.05(2) & \bf 0.98(1)\\ \hline
			
			\tabincell{c}{$\alpha=3$} &  $k_1=0$, $k_2=0$ & \bf 0.98(2) &  \bf 0.98(1) & \bf 0.98(2) \\ \hline\hline
			
		\end{tabular}
		\begin{tablenotes}
			\footnotesize
			\item[{[1]}] Those values of $F_{ij}$ in bold fonts are expected to equal one. All errors shown are one standard deviation with 0.7 million averages.
		\end{tablenotes}
		
		\label{Fidelity}
	\end{threeparttable}
\end{table*}
\section{Conclusions}
We have experimentally investigated the NH non-Abelian topological transition at the atomic scale. 
We observe the change in braid invariant and the corresponding relative phases, which successfully characterize this non-Abelian topological transition. 
The total charge remains unchanged and fails to characterize this transition. 
Furthermore, we demonstrate an intriguing process involving creation and coalescence of EPs. 
To ensure concise presentation, the method adopted for obtaining eigenvalues is specific to the model considered in this work. 
In future studies, one can prepare different initial states and perform multi-basis measurements on the final states evolved under NH Hamiltonians, which will ultimately enable prior-knowledge-free determination of the Hamiltonian's spectral properties (Supplementary Materials, section 4). 
Compared to previous work\cite{HarrisNature} that demonstrated non-commuting braids, our work on the NH non-Abelian topological transition highlights the dynamical interplay between multiple types of EPs. 
A key feature is that the properties of EPs may change during such processes. 
The two EPs $X,Y$, when considered together, undergo a shift in their properties due to the interplay with EPs $U,V$, leading to the formation of an EP3. 
The non-Abelian topology, arising from the multi-gap nature of complex NH bands, diverges from the stable topology governed by K-theory\cite{A,B,C}. 
In Hermitian systems, non-Abelian topology might also become relevant under multi-gap conditions, such as in nodal-line metals with specific symmetry constraints\cite{nattH1,nattH3,F} or in quantum dynamics\cite{nattH5,nattH6}.

The non-Abelian topological transitions explored in our work are expected to offer valuable insights into understanding the NH topological phenomena. 
For example, topologically protected band crossings in 2D lattice models\cite{closegap1} can evolve from gapped bands through such transitions. 
The NH skin effects\cite{skin} can be tailored through these transitions. 
Considering that the NV centre is a single-spin system, we can manipulate the spectral topology for single-spin systems to realize state control methods beyond the Abelian scenarios. 
For example, we can leverage the NH non-Abelian topological transition to create various high-order degeneracies\cite{highorderep1} and realize robust quantum control methods for qudit systems\cite{DDEP}. 
Our work also establishes a platform capable of exploring topological properties in atomic-scale systems. 
For instance, by including quantum jumps in the real time dynamics, we have the opportunities to observe exotic NH phenomena\cite{huyingprx,yiweiarxiv,andreiarxiv}.
 
\renewcommand{\thefigure}{S\arabic{figure}}
\renewcommand{\thetable}{S\arabic{table}}
\setcounter{figure}{0}
\section{methods}

	\subsection{Experiment setup}
	
	\quad \ \ The experiments were performed based on a setup for optically detected magnetic resonance. Green laser pulses of 532 nm were controlled using an acousto-optic modulator (ISOMET), passing through it twice before being directed through an oil objective (Olympus, UPLXAPO 100*O, NA 1.45). The emitted phonon sideband fluorescence, within the 650-800 nm wavelength range, was captured using the same oil objective and then detected by an avalanche photodiode (Perkin Elmer, SPCM-AQRH-14) with a counter card. A static magnetic field of 501 G, aligned with the NV centre's symmetry axis, was established using a permanent magnet. To manipulate the spin states of NV centre, an arbitrary waveform generator (Keysight, M8190A) produced microwave (MW), electric field (EF), and radio-frequency (RF) pulses. These pulses were then separately amplified by power amplifiers (two Mini Circuits ZHL-15W-422-S+ for the MW and EF pulses, and an LZY-22+ for the RF pulses).
	
	\subsection{Experimental sequence}
\begin{figure}[http]
		\vspace{-0em}
		\centering
		\includegraphics[width=1\columnwidth]{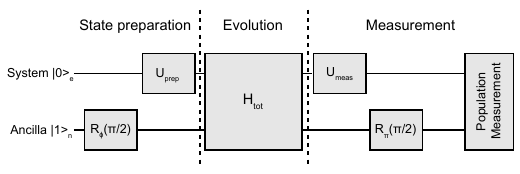}
		\caption{Schematic of the experimental sequence. The experimental realization contains three parts: the preparation, evolution under $H_tot$ and population measurements. The operations are implemented by microwave, electric field and radio-frequency pulses.
		}
		\label{ExtFig1}
	\end{figure}

	\quad \ \ The whole experimental sequence can be divided into three parts (Fig.\ref{ExtFig1}): the state preparation, NH evolution and population measurement. 
	First, the initial state of the form $U_{\rm prep}|0\rangle_e\otimes(|-\rangle_n+\eta_0|+\rangle_n)$ (not normalized for convenience) is prepared. Here, $\eta_0$ is a properly chosen number for experimental convenience, and $|\pm\rangle_n$ are the eigenstates of $\sigma_y$. In our experiment $\eta_0 = \sqrt{0.3}$ is chosen. The static magnetic field is set to 501 Gauss and optical pumping is applied to polarize the NV centre to the state $|0\rangle_e|1\rangle_n$. 
	Following the polarization stage, the single-qubit rotation $R_{\phi}(\pi/2)$ is applied on the nuclear spin. Here we adopt the convention $R_{\phi_0}(\theta_0)=e^{-iS_{\phi_0}\theta_0},S_{\phi_0}=\cos\phi_0S_x+\sin\phi_0S_y$, where $S_{x,y}$ are spin 1 or spin 1/2 operators for the electron or nuclear. The parameter $\phi$ is determined by $\phi = \left\lbrace \text{atan}\left[ 2\eta_0/(\eta_0^2-1) \right]+\pi \right\rbrace $, corresponding to the choice of $\eta_0$. Subsequently, the operation $U_{\rm prep}$ is performed on the electron spin to finish the state preparation. In our experiment, $U_{\rm prep}$ is chosen as $R^{ij}_{\chi}(\pi/2)=e^{-i(\sigma^{ij}_x\cos\chi+\sigma^{ij}_y\sin\chi)\pi/4}$, where $\chi$ and $(i,j)=(0,\pm1)$ (which means the effective qubit formed by $\ket{m_S=0}$ and $\ket{m_S=\pm1}$) depend on different experiment points to ensure that the population in target subspace is high enough to be measured.
	
	Second, the NH evolution is implemented using the time-dependent microwave and electric field pulses according to the dilation method (the details of amplitudes, frequencies and phases of the control pulses can be found in Supplementary materials, section 1). The whole state become $|\psi_i\rangle_e|-\rangle_n+\eta(t)|\psi_i\rangle_e|+\rangle_n$, where $|\psi_i\rangle_e$ is an eigenstate of $H$ (see Supplementary materials, section 5, for detail). We have utilized the assumption that the evolution time is long enough so that the desired eigenstate dominates. 
	
	Third, the operation $U_{\rm meas}$ is applied to the electron spin to change the measurement basis. Subsequently, the nuclear spin rotation $R_{\pi}(\pi/2)$ was applied to transform the state from $U_{\rm meas}|\psi_i\rangle_e |-\rangle_n + U_{\rm meas}\eta(t)|\psi_i\rangle_e |+\rangle_n$ to $U_{\rm meas}|\psi_i\rangle_e |1\rangle_n + U_{\rm meas}\eta(t)|\psi_i\rangle_e |0\rangle_n$ for the convenience of measurements. Finally, population measurement was performed on this state where the six populations $P_{\ket{m_S}_e\ket{m_I}_n}$ were obtained (see Supplementary materials, section 2).
	
	\subsection{Acquisition of the eigenvalues of the NH Hamiltonian}
	\quad \ \ The eigenvalues are determined from a set of equations specific to the model given by Eq.\ref{Model}. The details are presented as follows. Since the characteristic polynomial can also be written as 
	\begin{equation}
		P(E) = (E-E_1)(E-E_2)(E-E_3)= E^3 + c_2E^2 + (-c_1^2-2)E -c_1^2c_2,
	\end{equation} 
	we have 
	\begin{equation}
		\begin{aligned}
			E_1+E_2+E_3&=-c_2,\\
			E_1E_2+E_2E_3+E_1E_3&=-c_1^2-2,\\
			E_1E_2E_3&=c_1^2c_2,
		\end{aligned}
		\label{cE}
	\end{equation}
	where $E_i$ is the eigenvalues of the Hamiltonian, $c_1 = \sqrt{2}(i(\alpha+1)/4-k_1)$ and $c_2=-ik_2+1-(\alpha-2)^2$.
	Thus the eigenvalues instead of the parameters $c_1,c_2$ can be regarded as variables of the Hamiltonian, with the constraint equation 
	\begin{equation}
		E_1E_2E_3 = (E_1+E_2+E_3)(E_1E_2+E_2E_3+E_1E_3+2).
		\label{Econs}
	\end{equation}
	It can be shown that the eigenstate corresponding to the eigenvalue $E_i$ takes the form (not normalized for convenience)
	\begin{equation}
		\ket{\psi_i} = (-1+(c_1+E_i)(c_2+E_i),-(c_1+E_i),1)^T.
	\end{equation}
	According to Eq.\ref{cE}, we can replace $c_1,c_2$ by $E_i$ as $c_2 = -E_1-E_2-E_3,c_1 = \sqrt{-E_1E_2E_3/(E_1+E_2+E_3)}$. The eigenstates are then parameterized by the eigenvalues only. Another four equations can be formulated to that relate the three eigenvalues to population information of the eigenstates, which can be obtained through population measurements. Define the unitary operations as
	\begin{equation}
		\begin{aligned}
			&U_1 = \left[\begin{array}{ccc}
				0 & -i & 0\\
				-i & 0 & 0\\
				0 & 0 & 1
			\end{array}\right],\\
			&U_2 = \left[\begin{array}{ccc}
				1 & 0 & 0\\
				0 & 1/\sqrt{2} & -i/\sqrt{2}\\
				0 & -i/\sqrt{2} & 1/\sqrt{2}
			\end{array}\right],\\
			&U_3 = \left[\begin{array}{ccc}
				1/\sqrt{2} & -i/\sqrt{2} & 0\\
				-i/\sqrt{2} & 1/\sqrt{2} & 0\\
				0 & 0 & 1
			\end{array}\right].\\
		\end{aligned}
	\end{equation}
	These operations can all be realized by microwave pulses. Denote $P^{a,i}_j,P^{b,i}_j$ as the populations after the unitary operations $U_a = U_2U_1, U_b = U_3U_2U_1$ applied on the eigenstate $\ket{\psi_i}$, respectively. Here $i = 1,2,3$ label the eigenstates corresponding to the eigenvalues $E_i$, and $j = 1,2,3$ label the populations on the three levels. We choose the four quantities to be measured as $P^{a,i_1}_2/P^{a,i_1}_1$, $P^{b,i_1}_2/P^{b,i_1}_1$, $P^{a,i_2}_2/P^{a,i_2}_1$ and $P^{b,i_2}_2/P^{b,i_2}_1$. ($i_{1,2}$ are chosen by the convenience of experimental realization.)
	The explicit form of the equations can be formulated as
	\begin{equation}
		\begin{aligned}
			\frac{P^{a,i_1}_2}{P^{a,i_1}_1}&=\frac{1}{2}|c_2+E_{i_1}|^2,\\
			\frac{P^{b,i_1}_2}{P^{b,i_1}_1}&=|\frac{i+\frac{1}{\sqrt{2}}(c_2+E_{i_1})}{1+\frac{i}{\sqrt{2}}(c_2+E_{i_1})}|^2.\\
			\frac{P^{a,i_2}_2}{P^{a,i_2}_1}&=\frac{1}{2}|c_2+E_{i_2}|^2,\\
			\frac{P^{b,i_2}_2}{P^{b,i_2}_1}&=|\frac{i+\frac{1}{\sqrt{2}}(c_2+E_{i_2})}{1+\frac{i}{\sqrt{2}}(c_2+E_{i_2})}|^2.
		\end{aligned}
		\label{eqmeas}
	\end{equation}
	Combining Eq.\ref{cE}, Eq.\ref{Econs} and Eq.\ref{eqmeas}, the three eigenvalues can be solved. 
	As examples, for the section at $\alpha=0.39$, at the parameter $\theta=11\pi/8$ and $\theta=13\pi/8$ on $\Gamma$, the solved eigenvalues are $(E_1,E_2,E_3) = [2.3(3)-1.0(2)i,0.5(2)-0.1(3)i,-1.2(2)-0.3(3)i]$ and $[2.3(1)-1.1(2)i,0.1(2)-0.6(3)i,-0.9(2)+0.2(3)i]$, where the theoretical predictions are $(2.3-0.9i,0.5+0.2i,-1.2-0.6i)$ and $(2.4-1.0i,0.2-0.6i,-1.0+0.3i)$, respectively. The error bars are one standard deviation with 600,000 averages obtained via Monte Carlo method.
	
	\subsection{Acquisition of braids}
	\quad \ \ The details on deriving the braids from the relative phases between eigenvalues are presented as follows.
	
	The relative phase is defined as $\phi_{ij}=-\arg(E_i-E_j)$. The band indices $i,j$ are determined based on the real parts of eigenvalues at the initial point of the path $\theta=0$. The eigenvalue $E_i$ for $\theta>0$ is defined by continuation from $\theta=0$. A crossing occurs whenever $Re(E_i)=Re(E_j)$, i.e., when $\phi_{ij}=\pm\pi/2$. If $\phi_{ij}=\pi/2$, then $Re(E_i)=Re(E_j)$ and $Im(E_i)<Im(E_j)$. We denote this crossing as $\tau_{ij}$, which indicates that the band labeled by $j$ crosses the band labeled by $i$ from above. The $\phi_{ij}=-\pi/2$ case is denoted by $\tau_{ji}$. According to the above definitions, the four crossings from left to right in Fig.3(d) are $\tau_{12},\tau_{13},\tau_{32}$ and $\tau_{12}$, respectively. 
	
	Note that the band labels in the braid operator $\sigma_{i(i+1)}$ have different meanings. The labels are assigned according to the real parts of eigenvalues at the specific $\theta$ (not by continuation from $\theta=0$). To obtain the braid invariant from the information of relative phase, we need to translate the band labels from $\tau$ to $\sigma$. The transformation rule can be established as follows. The key is to track how the labels evolve with respect to $\theta$. Before the first crossing occurs, these two labels are identical. If a crossing marked by $\tau_{i(i+1)}$ happens, then the two bands labeled by $i,i+1$ are permuted. Further crossings amount to actions of more permutation operators. Suppose there are $m$ crossings along the entire path. We have the transformation relation between the labels in $\sigma$ and $\tau$ ($N$ is number of bands)
	\begin{equation}
		\sigma:~(1,2,...,N)\leftrightarrow
		\tau:~s_ms_{m-1}...s_2s_1(1,2,...,N).
	\end{equation}
	
	Here the left-hand side refers to the labels for $\sigma_{ij}$ and the right-hand side to $\tau_{ij}$. $s_i$ is the permutation operator associated with the $i^{th}$ crossing. With this transformation relation, the mapping for the $(m+1)^{th}$ crossing from $\tau_{ij}$ to $\sigma_{kl}$ is: $k=Z_m^{-1}(i)$ and $l=Z_m^{-1}(j)$, where $Z_m=s_ms_{m-1}...s_2s_1$.
	
\begin{figure}[http]
		\vspace{-0em}
		\centering
		\includegraphics[width=1\columnwidth]{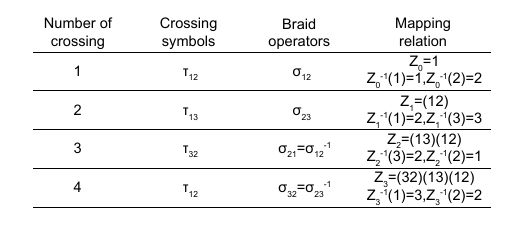}
		\caption{Transformation from $\tau$ to $\sigma$ for the crossings in Fig. 3(d). $\tau_{ij}$ represents the results obtained from measured relative phases, $\sigma_{ij}$ represents the corresponding braid operators, and $Z_i$ is the permutation operator that records the relation between $\tau_{ij}$ and $\sigma_{ij}$.
		}
		\label{ExtFig5}
	\end{figure}

	As an example, in Fig.\ref{ExtFig5} we list the translation from $\tau$ to $\sigma$ for the four crossings in Fig. 3(d). Here, we use cycle notation, e.g., $(ab)$, to denote the permutation operator that exchanges $a$ and $b$. We have also utilized the fact that $\sigma_{(i+1)i}=\sigma_{i(i+1)}^{-1}$. The permutation sequence in Fig. 3(d) then yields the braid invariant $b=\sigma_{12}\sigma_{23}\sigma_{12}^{-1}\sigma_{23}^{-1}$ as presented in the main text.

	This work was supported by the Innovation Program for Quantum Science and Technology (Grant No. 2021ZD0302200 (J.D.)), the National Natural Science Foundation of China (Grant Nos. T2388102 (X.R.), 12174373 (Yang Wu), 92265204 (Ya Wang), 12261160569 (X.R.) and 12474496 (H.H.)), National Key R\&D Program of China (Grant Nos. 2023YFA1406704 (H.H.) and 2022YFA1405800 (H.H.)), the Chinese Academy of Sciences (Grant Nos. XDC07000000 (J.D.) and GJJSTD20200001 (J.D.)) and Hefei Comprehensive National Science Center (J.D.). Ya W. and Yang W. thanks the Fundamental Research Funds for the Central Universities for their support. This work was partially carried out at the USTC Center for Micro and Nanoscale Research and Fabrication.

\renewcommand{\theequation}{S\arabic{equation}}
\renewcommand{\bibnumfmt}[1]{[RefS#1]}
\renewcommand{\citenumfont}[1]{RefS#1}	
\setcounter{equation}{0}
\setcounter{table}{0}

\begin{figure*}[http]
	\vspace{-0em}
	\centering
	\includegraphics[width=2\columnwidth]{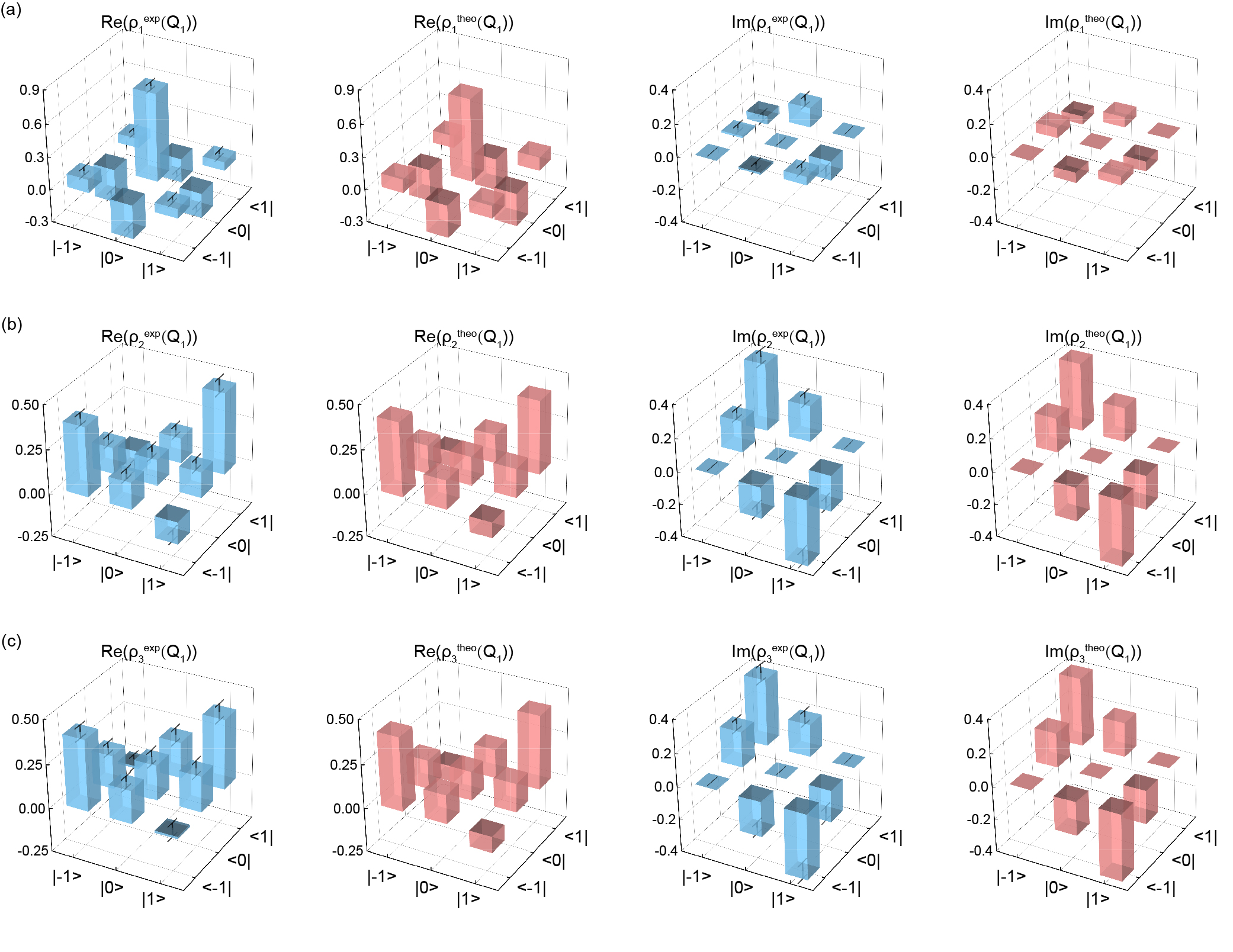}
	\caption{Eigenstates of the non-Hermitian Hamiltonian corresponding to the EP2 at $\alpha=0.39,k_1=k_2=0$. (a-c), Real and imaginary parts of the measured density matrices $\rho_1exp(Q_1)$ (a), $\rho_2exp(Q_1)$ (b) and $\rho_3exp(Q_1)$ (c) of three eigenstates (labeled by 1,2 and 3) obtained by quantum state tomography, with corresponding theoretical predictions $\rho_{1,2,3}exp(Q_1)$. $Q_1$ marks the EP2 at $\alpha=0.39,k_1=k_2=0$. All errors shown are one standard deviation with 0.7 million averages.
	}
	\label{ExtFig2}
\end{figure*}
\begin{figure*}[http]
	\vspace{-0em}
	\centering
	\includegraphics[width=2\columnwidth]{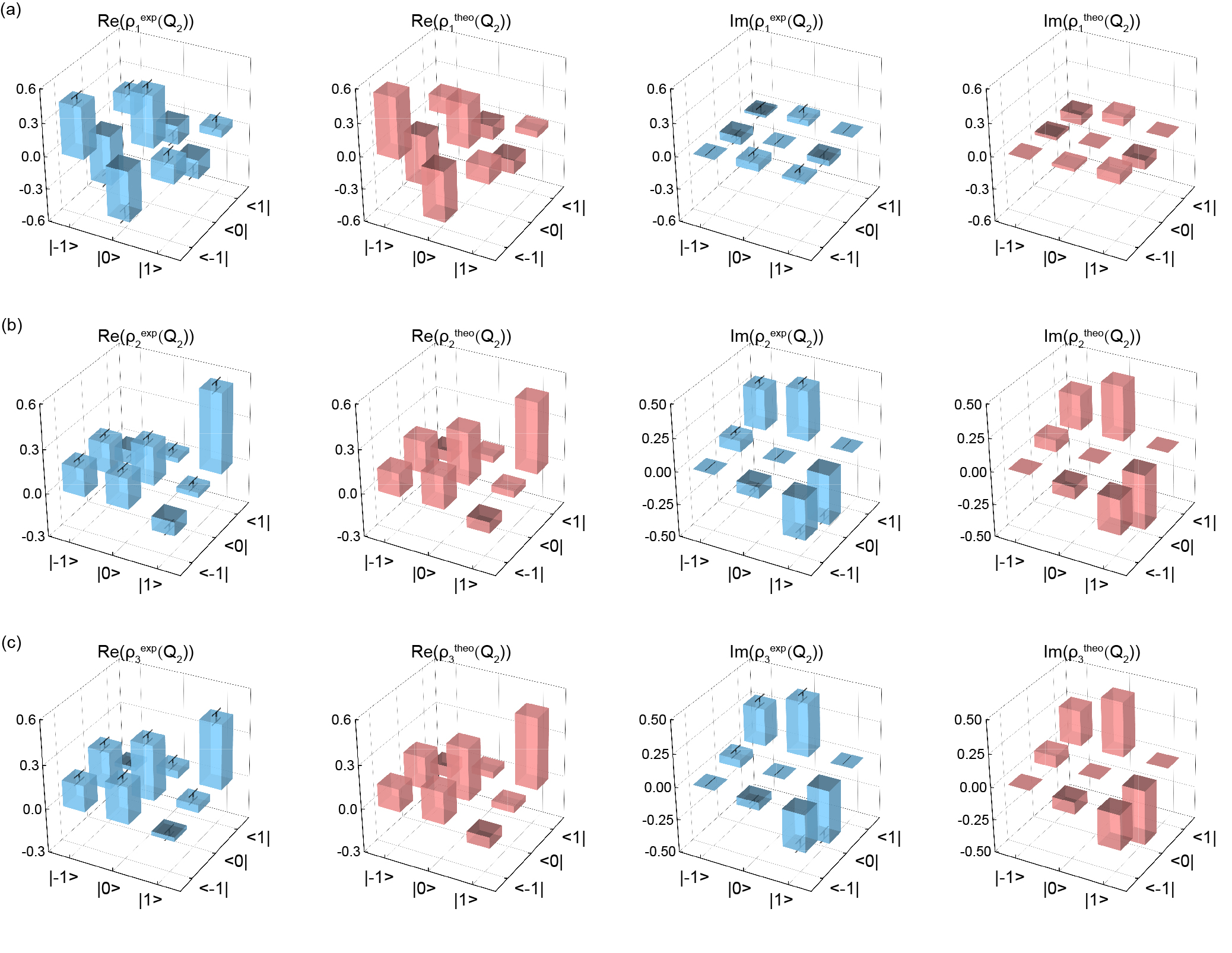}
	\caption{Eigenstates of the non-Hermitian Hamiltonian corresponding to the EP2 at $\alpha=1,k_1=0.46,k_2=-1.06$. (a-c), Real and imaginary parts of the measured density matrices $\rho_1exp(Q_2)$ (a), $\rho_2exp(Q_2)$ (b) and $\rho_3exp(Q_2)$ (c) of three eigenstates (labeled by 1,2 and 3) obtained by quantum state tomography, with corresponding theoretical predictions $\rho_{1,2,3}exp(Q_2)$. $Q_2$ marks the EP2 at $\alpha=1,k_1=0.46,k_2=-1.06$. All errors shown are one standard deviation with 0.7 million averages.
	}
	\label{ExtFig3}
\end{figure*}
\begin{figure*}[http]
	\vspace{-0em}
	\centering
	\includegraphics[width=2\columnwidth]{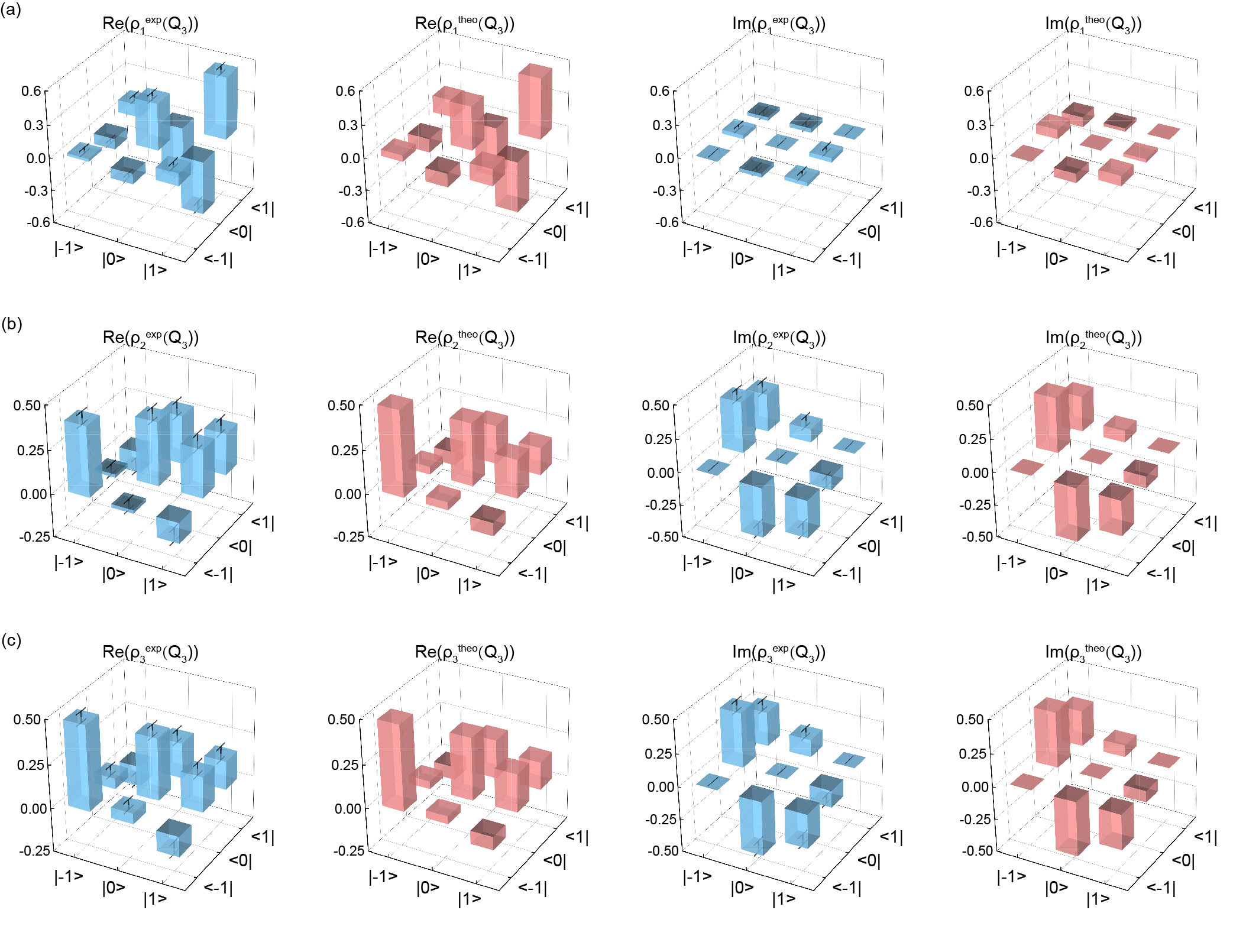}
	\caption{Eigenstates of the non-Hermitian Hamiltonian corresponding to the EP2 at $\alpha=1,k_1=-0.46,k_2=1.06$. (a-c), Real and imaginary parts of the measured density matrices $\rho_1exp(Q_3)$ (a), $\rho_2exp(Q_3)$ (b) and $\rho_3exp(Q_3)$ (c) of three eigenstates (labeled by 1,2 and 3) obtained by quantum state tomography, with corresponding theoretical predictions $\rho_{1,2,3}exp(Q_3)$. $Q_3$ marks the EP2 at $\alpha=1,k_1=-0.46,k_2=1.06$. All errors shown are one standard deviation with 0.7 million averages.
	}
	\label{ExtFig4}
\end{figure*}

\clearpage	
	\onecolumngrid
	\vspace{1.5cm}
	\begin{center}
		\textbf{\large Supplementary Material}
	\end{center}
	
	\setcounter{figure}{0}
	\setcounter{equation}{0}
	\setcounter{table}{0}
	\makeatletter

	

\section{Dilation method and the design of pulse parameters}
In this section we show details of designing the pulses to realize the dilated Hamiltonian given by the dilation method. 

Based on the dilation method, the NH Hamiltonian is embedded into the dilated Hamiltonian taking the form
\begin{equation}
	H_{\rm tot}(t)=\Xi(t)\otimes\ket{1}\bra{1}+\Lambda(t)\otimes\ket{0}\bra{0},
\end{equation}
where $\Xi(t)$ and $\Lambda(t)$ are Hermitian operators on the system, $\ket{0,1}$ are states of the ancilla qubit.
The form of $\Xi(t)$ and $\Lambda(t)$ depends on the NH Hamiltonian $H$. Explicitly, $\Xi=\hat{\Lambda}+\hat{\Xi},\Lambda=\hat{\Lambda}-\hat{\Xi}$, where 
\begin{equation}
	\hat{\Lambda}=[H+(i\partial_t\eta+\eta H)\eta] M^{-1}, \hat{\Xi} = i( H\eta-\eta H-i\partial_t\eta) M^{-1}.
\end{equation} 
For simplicity, we have dropped the $t$ dependence. The explicit forms of $M$ and $\eta$ are 
\begin{equation}
	M(t) = e^{-iH^\dagger t}M(0)e^{iHt},\eta(t)=U(t)[M(t)-I]^{1/2}.
\end{equation}
The choice of $M(0)$ and $U(t)$ is flexible. The details of the derivations can be found in Ref.\cite{Dilation}. In our experiment, $M(0)=1.3I$ is chosen. In order to facilitate the realization of $H_{\rm tot}$ in NV center, $\hat{H}_{\rm tot} = s H_{\rm tot}$ is implemented, where $s$ is a nonzero coefficient that scales the evolution time. This is equivalent to apply the dilation method to the NH Hamiltonian $sH$ instead of $H$ (in our experiment, $s$ is of the order of 20-40 kHz, depending on different points). 

Then we focus on the design of pulse parameters for implementing $H_{\rm tot}$ in NV center. $H_{\rm tot}$ is realized in NV center by implementing a control Hamiltonian and choosing an appropriate interaction picture. The control Hamiltonian takes the following form (the integral variable $\tau$ is omitted)
\begin{equation}
	\begin{aligned}
		H_C(t) &= 2\sqrt{2}\pi\Omega_{\rm MW2} \cos[\int_{0}^{t}\omega_{\rm MW2}+\phi_{\rm MW2}(t)]S_x\otimes\ket{1}\bra{1}\\
		&+2\sqrt{2}\pi\Omega_{\rm MW1}  \cos[\int_{0}^{t}\omega_{\rm MW1}+\phi_{\rm MW1}(t)]S_x\otimes\ket{1}\bra{1}\\
		&+2\pi\Omega_{\rm EF1}  \cos[\int_{0}^{t}\omega_{\rm EF1}+\phi_{\rm EF1}(t)]S_{13}\otimes\ket{1}\bra{1}\\
		&+2\sqrt{2}\pi\Omega_{\rm MW4}  \cos[\int_{0}^{t}\omega_{\rm MW4}+\phi_{\rm MW4}(t)]S_x\otimes\ket{0}\bra{0}\\
		&+2\sqrt{2}\pi\Omega_{\rm MW3}  \cos[\int_{0}^{t}\omega_{\rm MW3}+\phi_{\rm MW3}(t)]S_x\otimes\ket{0}\bra{0}\\
		&+2\pi\Omega_{\rm EF2}  \cos[\int_{0}^{t}\omega_{\rm EF2}+\phi_{\rm EF2}(t)]S_{13}\otimes\ket{0}\bra{0},
	\end{aligned}
\end{equation}
where 
\begin{equation}
	S_x = \frac{1}{\sqrt{2}}\left[\begin{array}{ccc}
		0 & 1 & 0\\
		1 & 0 & 1\\
		0 & 1 & 0
	\end{array}\right], S_{13} = \left[\begin{array}{ccc}
		0 & 0 & 1\\
		0 & 0 & 0\\
		1 & 0 & 0
	\end{array}\right].
\end{equation}
The parameters to be determined are $\Omega_{\zeta}$, $\phi_{\zeta}$ and $\omega_{\zeta}$, where $\zeta\in\lbrace\rm MW1,MW2,MW3,MW4,EF1,EF2\rbrace$.
The interaction picture is chosen as
\begin{equation}
	U_{\rm rot} = \exp[i\int_{0}^{t}H_{\rm NV}-\text{diag}(d_1,d_2,\dots,d_6)],
\end{equation}
where $H_{\rm NV}$ is the Hamiltonian for the ground state of NV center and $d_1,d_2,\dots,d_6$ are the diagonal elements of $H_{\rm tot}$.
The Hamiltonian under the rotating wave approximation can be written as
\begin{equation}
	\begin{aligned}
		&H_{\rm rot}=U_{\rm rot}H_CU_{\rm rot}^\dagger+\text{diag}(d_1,d_2,\dots,d_6)\\
		&=\left[\begin{array}{ccc}
			d_1 & A_1 & C_1\\
			A_1^* & d_2 & B_1\\
			C_1^* & B_1^* & d_3
		\end{array}\right]\otimes\ket{1}\bra{1}+\left[\begin{array}{ccc}
			d_4 & A_2 & C_2\\
			A_2^* & d_5 & B_2\\
			C_2^* & B_2^* & d_6
		\end{array}\right]\otimes\ket{0}\bra{0},
	\end{aligned}
\end{equation}
where
\begin{equation}
	\begin{aligned}
		&A_1=\pi\Omega_{\rm MW2} e^{-i\phi_{\rm MW2}-i(\int_{0}^{t}\omega_{\rm MW2}-\omega_{12}-d_2+d_1)},\\
		&B_1=\pi\Omega_{\rm MW1} e^{i\phi_{\rm MW1}+i(\int_{0}^{t}\omega_{\rm MW1}-\omega_{23}-d_2+d_3)},\\
		&C_1=\pi\Omega_{\rm EF1} e^{-i\phi_{\rm EF1}-i(\int_{0}^{t}\omega_{\rm EF1}-\omega_{13}-d_3+d_1)},\\
		&A_2=\pi\Omega_{\rm MW4} e^{-i\phi_{\rm MW4}-i(\int_{0}^{t}\omega_{\rm MW4}-\omega_{45}-d_5+d_4)},\\
		&B_2=\pi\Omega_{\rm MW3} e^{i\phi_{\rm MW3}+i(\int_{0}^{t}\omega_{\rm MW3}-\omega_{56}-d_5+d_6)},\\
		&C_2=\pi\Omega_{\rm EF2} e^{-i\phi_{\rm EF2}-i(\int_{0}^{t}\omega_{\rm EF2}-\omega_{46}-d_6+d_4)}.\\
	\end{aligned}
\end{equation}
Here we label the energy levels $\ket{m_S=1,0,-1}_e\otimes\ket{m_I=1}_n$ ($\ket{m_S=1,0,-1}_e\otimes\ket{m_I=0}_n$) as 1, 2 and 3 (4, 5 and 6) for simplicity and $\omega_{ij}$ $(\omega_{ij}>0)$ is the transition frequency between the levels $i$ and $j$.
Comparing $H_{\rm rot}$ and $H_{\rm tot}$, all the parameters $\Omega_{\zeta}$, $\phi_{\zeta}$ and $\omega_{\zeta}$ can be solved.

\section{Acquisition of the population information}
In this section we introduce the normalization procedures in our experiment and the acquisition of populations from the experimental data. In this section, $L_i$ ($i=1,2,...,6$) stands for the photoluminescence rate for the level labeled by $i$, $L$ is the column vector with its elements equal to $L_i$, and $\xi$ is the polarization for the electron spin. The $\pi$ rotations between the level $i$ and $j$ are denoted as $R_{ij}(\pi)$.

Since the electron spin state is not ideally polarized after the initial laser polarization procedure, we first characterize the polarization of the electron spin. The polarization of the electron spin is measured by a set of independent normalization sequences. Four different sequences are chosen (after the polarization by optical pumping) to obtain a set of equations for $L_i$ and $\xi$. The four sequences are $\mathbb{I}$, $R_{23}(\pi)$, $R_{25}(\pi)$, and $R_{23}(\pi)+R_{25}(\pi)$. Here $\mathbb{I}$ means the identity operation, i.e., no pulse is applied. The set of equations relating $\xi$ and measured counts is formulated as
\begin{equation}
	\begin{aligned}
		&L_1(1-\xi)/2+L_2\xi+L_3(1-\xi)/2 = C^{\rm Pol}_1,\\
		&L_1(1-\xi)/2+L_3\xi+L_2(1-\xi)/2 = C^{\rm Pol}_2,\\
		&L_1(1-\xi)/2+L_5\xi+L_3(1-\xi)/2 = C^{\rm Pol}_3,\\
		&L_1(1-\xi)/2+L_3\xi+L_5(1-\xi)/2 = C^{\rm Pol}_4,\\
	\end{aligned}
\end{equation}
where $C^{\rm Pol}_j$ is the measured counts for sequence $j$. The polarization $\xi$ can be solved from $(1-\xi)/2\xi = (C^{\rm Pol}_2-C^{\rm Pol}_4)/(C^{\rm Pol}_1-C^{\rm Pol}_3)$. In our experiment, the result is $\xi=0.98(1)$. 

In order to extract the population information in the system subspace, both the normalization and measurement sequences are implemented. 
The normalization sequences are utilized to obtain the photoluminescence rate $L_i$ ($i=1,2,...,6$) for each level. After the polarization stage, the state of the total system was $\rho_{\rm ini} = (1-\xi)/2\ket{1}\bra{1}+\xi\ket{2}\bra{2}+(1-\xi)/2\ket{3}\bra{3}$. 
By combining different MW and RF $\pi$ pulses, $\rho_{\rm ini}$ is transformed to the form $\rho = (1-\xi)/2\ket{i}\bra{i}+\xi\ket{j}\bra{j}+(1-\xi)/2\ket{k}\bra{k}$ ($i\ne j,j\ne k,i\ne k, 1\le i,j,k\le 6$). The photoluminescence rate for this state is then measured. Here we used six different normalization sequences to obtain $L$. The set of equations for $L$ can be formulated as
\begin{equation}
	\begin{aligned}
		&L_1(1-\xi)/2+L_2\xi+L_3(1-\xi)/2 = C^{\rm norm}_1,\\
		&L_2(1-\xi)/2+L_1\xi+L_3(1-\xi)/2 = C^{\rm norm}_2,\\
		&L_1(1-\xi)/2+L_3\xi+L_2(1-\xi)/2 = C^{\rm norm}_3,\\
		&L_1(1-\xi)/2+L_6\xi+L_2(1-\xi)/2 = C^{\rm norm}_4,\\
		&L_1(1-\xi)/2+L_5\xi+L_3(1-\xi)/2 = C^{\rm norm}_5,\\
		&L_1(1-\xi)/2+L_4\xi+L_3(1-\xi)/2 = C^{\rm norm}_6,\\
	\end{aligned}
\end{equation}
where $C_l^{\rm norm}$ is the measured counts for each normalization sequences. Here the polarization $\xi$ is taken as the value obtained by the independent polarization measurement. Finally by solving these equations, $L_i$ can be obtained. 

The measurement sequences are utilized to solve the population of each level in our experiment. Knowing the photoluminescence rate of each level, the populations can be obtained using a similar method. Denote $p$ as the column vector recording the populations of the six levels, $M_m$ as a permutation matrix depending on the combination of the $\pi$ pulses and $C_m^{\rm exp}$ as the measured counts. The equations for different measurement sequences are $L^T M_m p = C_m^{\rm exp}$ ($m = 1,2,...,5$), where five different measurement sequences are performed. The explicit form of the equations in our experiment is
\begin{equation}
	\begin{aligned}
		&L_1p_1+L_2p_2+L_3p_3+L_4p_4+L_5p_5+L_6p_6 = C^{\rm exp}_1,\\
		&L_1p_2+L_2p_1+L_3p_3+L_4p_4+L_5p_5+L_6p_6 = C^{\rm exp}_2,\\
		&L_1p_1+L_2p_3+L_3p_2+L_4p_4+L_5p_5+L_6p_6 = C^{\rm exp}_3,\\
		&L_1p_1+L_2p_6+L_3p_2+L_4p_4+L_5p_5+L_6p_3 = C^{\rm exp}_4,\\
		&L_1p_1+L_2p_5+L_3p_3+L_4p_4+L_5p_2+L_6p_6 = C^{\rm exp}_5,\\
		&p_1+p_2+p_3+p_4+p_5+p_6 = 1.
	\end{aligned}
	\label{Eqp}
\end{equation}
The populations of each level are solved from Eq.\ref{Eqp}. 
The population information in the target subspace spanned by $\ket{1,2,3}$ is reconstructed as $P_{\ket{m_S=1,0,-1}_e} = p_{1,2,3}/(p_1+p_2+p_3)$.

\section{Positions of EPs of the model Hamiltonian}
The characteristic polynomial of the Hamiltonian in Eq.1 in the main text is
\begin{equation}
	P(E) = E^3 + c_2E^2 + (-c_1^2-2)E -c_1^2c_2, 
	\label{poly}
\end{equation}
where $c_1 = \sqrt{2}(i(\alpha+1)/4-k_1), c_2=-ik_2+1-(\alpha-2)^2$. The eigenvalues $E_i$ are roots of the equation $P(E)=0$. The positions of EPs can be obtained by finding the degenerate points of eigenvalues. The polynomial equation $P(E)=0$ has multiple roots when the corresponding discriminant is zero. The discriminant of $P(E)$ takes the form
\begin{equation}
	\Delta_\alpha(k_1,k_2)=c_2^2(c_1^2+2)^2+4(c_1^2+2)^3+4c_2^4c_1^2-27c_1^4c_2^2+18c_2^2c_1^2(c_1^2+2),
\end{equation}
where $c_1,c_2$ depend on $\alpha,k_1$ and $k_2$.
The position of EPs can be found by solving the equation $\Delta_\alpha(k_1,k_2)=0$.

\section{Generalized method of acquisition of eigenvalues for general Hamiltonians}

In this section, we provide a generalized method that can be applied to obtain the eigenvalue information of general 3 by 3 Hamiltonians. The main idea is to extract all parameters of the Hamiltonian and then obtain its eigenvalues from state evolution under the Hamiltonian. 

For an arbitrary 3 by 3 non-Hermitian Hamiltonian, it has 18 degrees of freedom. Since $H$ and $H-d\mathbb{I}$ have the same braiding behaviors, removing the trace part of the Hamiltonian does not affect the characterization of braids. We can then eliminate the freedom of the trace of the Hamiltonian. Thus the degrees of freedom of $H$ are reduced to $16$.

Denote the Hamiltonian as $H=H(\vec{a})$, where $\vec{a}$ represents the parameters and it has $16$ components. We label the three levels as $1,2,3$. In the experiment we have access to the quantities of the form $P_i/(P_1+P_2+P_3), i=1,2,3$. Here $P_i$ is the probability of finding the state in level $i$. By preparing the initial state as $\ket{\psi_{ini}}$, evolving under $H$ for time $t$, and measuring the population under the bases marked by $\ket{\psi_{pro,i}}$, the general form of $P_i$ is $|\bra{\psi_{pro,i}}e^{-iHt}\ket{\psi_{ini}}|^2$. The choices of $\ket{\psi_{ini}}$ and $\ket{\psi_{pro,i}}$ are flexible in the experiment, e.g., the eigenstates of $S_{x,y,z}$. For each choice of $\ket{\psi_{ini}}$ and $\ket{\psi_{pro,i}}$, we can construct a constraint equation that relates the parameters $\vec{a}$ and the measured quantity $M_k$ as
\begin{equation}
	M_k = F_k(\vec{a};t).
\end{equation}
Here $k$ labels different choices, and $F_k(\vec{a};t)=P_i/(P_1+P_2+P_3)$ with $P_i = |\bra{\psi_{pro,i}}e^{-iH(\vec{a})t}\ket{\psi_{ini}}|^2$.
By choosing different initial states and measurement bases, we can either solve for $\vec{a}$ from $16$ measured quantities, or varying the time $t$ and extract $\vec{a}$ from fitting. Knowing the parameters of the Hamiltonian (except for the trace part), the eigenvalues can be obtained. The braids can then be constructed as described in the main text.

\section{Acquisition of the eigenstates of the NH Hamiltonian}
The eigenstates of an NH Hamitonian $H$ can be obtained from the steady states under the evolution of the NH Hamiltonian $g(H)$, where $g(x)$ is an analytic function of $x$. We label the eigenvalues of $g(H)$ as $\epsilon_{1,2,3}=g(E_{1,2,3})$ with the corresponding eigenstates as $\ket{\psi_{1,2,3}}$. Then for any initial state $\ket{\psi_{\rm ini}} = c_1\ket{\psi_1}+c_2\ket{\psi_2}+c_3\ket{\psi_3}$ ($c_1c_2c_3\ne0$), the evolution governed by $g(H)$ gives
\begin{equation}
	\ket{\psi} = c_1e^{-i\epsilon_{1}t}\ket{\psi_1}+c_2e^{-i\epsilon_{2}t}\ket{\psi_2}+c_3e^{-i\epsilon_{3}t}\ket{\psi_3}.
\end{equation}
Thus when the evolution time $t$ is long enough, the state approaches the eigenstate corresponding to the eigenvalue with largest imaginary part. By changing the form of $g(H)$, all three eigenstates can be approached. With the help of dilation method, as well as the quantum state tomography applied on the subspace where the nuclear spin is $\ket{-}_n$, the density matrix of the eigenstates can be obtained. Since direct results from quantum state tomography may give an unphysical state, the maximum likelihood estimation method was employed to obtain physical results\cite{Tomography}. For the model Hamiltonian, we can choose from $g(H)=H,-H,iH,-iH$ to obtain desired eigenstates for most cases. For a very few cases, we used $g(H)=\pm i/(H-\mathcal{E}\mathbb{I})$, where $\mathcal{E}$ is a properly chosen complex number which ensures that the eigenstate corresponding the eigenvalue closest to $\mathcal{E}$ can be obtained. 
Apart from the measured density matrices corresponding to the eigenstates at $(k_1,k_2,\alpha)=(0,0,3)$ shown in the main text, the results for $(k_1,k_2,\alpha)=(0,0,0.39),(0.46,-1.06,1),(-0.46,1.06,1)$ are shown in the Fig.\ref{ExtFig2}\ref{ExtFig3}\ref{ExtFig4}. For simplicity, we mark these three points $(k_1,k_2,\alpha)=(0,0,0.39),(0.46,-1.06,1)$ and $(-0.46,1.06,1)$ as $Q_1,Q_2$ and $Q_3$. The errors bars are one standard deviation with 700,000 averages obtained via Monte Carlo method.

\section{Homotopical classification of the eigenvalue topology in non-Hermitian systems}

The classification method uses homotopy theory and is adapted from references\cite{A,B,C,D}. For a given non-Hermitian exceptional point, we choose a closed trajectory enclosing it. The non-Hermitian bands are separable along this trajectory. Our strategy is to identify the classifying space of non-Hermitian bands along the loop, with its fundamental group yielding the topological invariant associated with the
exceptional degeneracy. 

For an $N$-band non-Hermitian Hamiltonian, the classifying space of separable bands is $X_N=Conf_N(\mathbb{C})\times F_N/S_N$. Here, $Conf_N(\mathbb{C})$ refers to the space of ordered N-tuples of complex eigenvalues, $F_N=U(N)/U^N(1)$ refers to the space of eigenstates, and $S_N$ refers to the permutation group. The fundamental group of $F_N$ is $\pi_1(p,F_N)=0$ ($p$ is some reference point on the loop), which is trivial. Then $\pi_1(p,X_N)=\pi_1(p,Conf_N(\mathbb{C})/S_N)=B_N$ gives the braid group. Thus the exceptional degeneracy is characterized by $B_N$. The braid invariant arises from the complex nature of eigenvalues in non-Hermitian systems.

We note the difference from the non-Abelian topology in $PT,C_2T$ symmetric systems\cite{D1,E,F}. To obtain the classifying space, the eigenvalue part $Conf_N(\mathbb{C})$ should be replaced by $Conf_N(\mathbb{R})$ and $F_N/S_N$ is replaced by $O(N)/\mathbb{Z}_2^N$. Since the eigenvalues are real, the contribution from the eigenvalues is trivial. The nontriviality arises from the eigenstate sector. For the case of three bands, the multigap condition yields the quaternion charge as the topological invariant\cite{D1}.

\end{document}